\documentclass[twocolumn]{aastex62}
\usepackage{array} 
\usepackage{amsmath}
\usepackage{wasysym}
\usepackage[shortcuts]{extdash}
\usepackage{xcolor}

\graphicspath{{figs/}}

\newenvironment{tightcenter}{%
	\setlength\topsep{0pt}
	\setlength\parskip{0pt}
	\begin{center}
}{%
 	\end{center}
}

\newcommand{\angstrom}{\textup{\AA}}
\newcommand\chandra{{\it Chandra}}

\newcommand\ciao{CIAO}

\newcommand\caldb{CALDB}

\newcommand\alphaox{\alpha_{\rm ox}}
\newcommand\rl{\mathcal{R}}

\shorttitle{X-ray Properties of Quasars at $z>4.5$}
\shortauthors{Snios et al.}

\begin{document}

\title{X-ray Properties of Young Radio Quasars at $z>4.5$}

\author{Bradford Snios} 
\affil{Harvard-Smithsonian Center for Astrophysics, 60 Garden Street, Cambridge, MA 02138, USA}
\author{Aneta Siemiginowska} 
\affil{Harvard-Smithsonian Center for Astrophysics, 60 Garden Street, Cambridge, MA 02138, USA}
\author{Ma{\l}gosia Sobolewska}
\affil{Harvard-Smithsonian Center for Astrophysics, 60 Garden Street, Cambridge, MA 02138, USA}
\author{C. C. Cheung}
\affil{Space Science Division, Naval Research Laboratory, Washington, DC 20375, USA}
\author{Vinay Kashyap}
\affil{Harvard-Smithsonian Center for Astrophysics, 60 Garden Street, Cambridge, MA 02138, USA}
\author{Giulia Migliori}
\affil{Department of Physics and Astronomy, University of Bologna, Via Gobetti 93/2, 40129 Bologna, Italy}
\affil{INAF--Institute of Radio Astronomy, Bologna, Via Gobetti 101, I-40129 Bologna, Italy}
\author{Daniel A. Schwartz} 
\affil{Harvard-Smithsonian Center for Astrophysics, 60 Garden Street, Cambridge, MA 02138, USA}
\author{{\L}ukasz Stawarz} 
\affil{Astronomical Observatory of the Jagiellonian University, ul. Orla 171, 30-244 Krak{\'o}w, Poland}
\author{Diana M. Worrall}
\affil{H. H. Wills Physics Laboratory, University of Bristol, Tyndall Ave, Bristol BS8 1TL, UK}

\begin{abstract}
We present a comprehensive analysis of \chandra\ X-ray observations of 15 young radio quasars at redshifts $4.5 < z < 5.0$. All sources are detected in the \mbox{0.5--7.0\,keV} energy band. Emission spectra are extracted, and the average photon index for the sample is measured to be $1.5\pm0.1$. Unabsorbed rest-frame 2--10\,keV luminosities are found to range between $(0.5$--$23.2) \times 10^{45}\rm\,erg\,s^{-1}$. The optical--X-ray  power-law spectral index $\alphaox$ is calculated for each source using optical/UV data available in the literature. The $\alphaox$--UV relationship is compared with other quasar surveys, and an anticorrelation is observed that agrees with independent estimates. Rest-frame radio and X-ray luminosities are established for the sample, and a correlation between the luminosities is detected. These multiwavelength results reinforce a lack of spectral evolution for quasars over a broad redshift range. We additionally identify three quasars from our multiwavelength analysis that are statistically significant outliers, with one source being a Compton-thick candidate in the early universe, and discuss each in detail. 
\end{abstract}


\section{Introduction}
\label{sect:intro}

The number of known quasars at redshift $z>4$ has drastically increased in recent years \citep[e.g.,][]{Banados2018}. This quasar population is powered by supermassive black holes (SMBHs) of $\sim$\,$10^{8}$--$10^{10} M_{\odot}$ \citep{Shlosman1990}, where large-scale jets and lobes are known to accompany radio quasars \citep{Begelman1984,Bridle1984}. X-ray intensity from quasars is dominated by emission from the innermost region of an accreting SMBH, such as the corona or the base of a jet \citep{Begelman1983}. A minor fraction of the total X-rays can also be generated from interactions between quasar radio jets and the interstellar medium (ISM) as strong shocks are driven into the ISM, heating and ionizing the gas \citep{Scheuer1974,Begelman1984}. 

In the case of young radio quasars, the radio source is expected to be contained within the ISM of the host galaxy on distance scales of \mbox{$\sim$\,0.01--10\,kpc}. The initial expansion of the shocks by the young radio jets is supersonic, and the radio lobes are highly overpressured with respect to their surroundings. Elevated thermal radiation is therefore predicted to contribute appreciably to the spatially extended X-ray intensity from these young radio quasars \citep{Heinz1998,Reynolds2001,Stawarz2008b,Mukherjee2018}. Hence, X-ray observations of young radio quasars offer a unique insight into the initial phases of feedback processes between the active galactic nuclei (AGNs) and its environment.

\begin{table*}
	\caption{\chandra{} Observations of $4.5 < z < 5.0$ Quasar Sample}
	\begin{tightcenter}
	\label{table:obs}
	\begin{tabular}{ c c c c c c c c c }
		\hline \hline
		Object & $z$\tablenotemark{a} & R.A.\tablenotemark{b} & Decl.\tablenotemark{b} & ObsID\tablenotemark{c} & Observation & $t_{\rm exp}$\tablenotemark{d} & Ref.\tablenotemark{e} \\
        & & (J2000) & (J2000) & & Date & (ks) & \\
		\hline
        J0311$+$0507 & 4.51 & 03:11:47.97 & +05:08:03.9 & 20475 & 2018 Aug 08 & 5.99 & --- \\ 
        J0813$+$3508 & 4.92 & 08:13:33.33 & +25:08:10.8 & 18444 & 2015 Dec 19 & 5.99 & 1 \\
        J0940$+$0526 & 4.50 & 09:40:04.80 & +05:26:30.9 & 20476 & 2017 Dec 31 & 5.99 & --- \\
        J1013$+$2811 & 4.75 & 10:13:35.44 & +28:11:19.2 & 20477 & 2018 Jan 22 & 5.99 & --- \\
        J1235$-$0003 & 4.69 & 12:35:03.05 & $-$00:03:31.8 & 20478 & 2018 Feb 26 & 5.99 & --- \\
        J1242$+$5422 & 4.73 & 12:42:30.59 & +54:22:57.5 & 18447 & 2016 May 16 & 4.87 & 1 \\
        J1311$+$2227 & 4.61 & 13:11:21.32 & +22:27:38.6 & 20479 & 2018 Mar 14 & 5.99 & --- \\
		J1400$+$3149 & 4.64 & 14:00:25.42 & +31:49:10.7 & 20480 & 2018 Feb 07 & 5.99 & --- \\
        J1454$+$1109 & 4.93 & 14:54:59.31 & +11:09:27.9 & 20481 & 2018 Jan 06 & 5.99 & --- \\
        J1548$+$3335 & 4.68 & 15:48:24.01 & +33:35:00.1 & 20482 & 2018 Feb 11 & 6.11 & --- \\
        J1606$+$3124 & 4.56 & 16:06:08.52 & +31:24:46.5 & 20483 & 2018 Feb 03 & 6.34 & --- \\
        J1611$+$0844 & 4.55 & 16:11:05.65 & +08:44:35.5 & 20484 & 2018 Jan 14 & 5.99 & --- \\
        J1628$+$1154 & 4.47 & 16:28:30.47 & +11:54:03.5 & 20485 & 2018 Feb 09 & 5.99 & --- \\
        J1659$+$2101 & 4.78 & 16:59:13.23 & +21:01:15.8 & 20486 & 2017 Dec 26 & 5.99 & --- \\
        J2102$+$6015 & 4.58 & 21:02:40.22 & +60:15:09.8 & 20487 & 2018 Mar 02 & 5.99 & --- \\
		\hline
	\end{tabular}
	\end{tightcenter}
	\tablenotemark{a}{Redshift measurements from \cite{Titov2013} and \cite{Paris2018}.}
	\tablenotemark{b}{Coordinates from VLBI positions reported in \cite{Coppejans2017}, and references therein.}
	\tablenotemark{c}{Observations performed using \chandra\ ACIS-S instrument with aimpoint on S3 chip.}
	\tablenotemark{d}{Total exposure time after flare removal reprocessing.}
	\tablenotemark{e}{References: (1) \cite{Zhu2019}.}
\end{table*}

X-ray studies of young radio quasars at low redshift ($z < 2$) indicate a complex diversity in emission over the 0.5--40\,keV rest-frame energy range \citep[e.g.,][]{Siemiginowska2008, Tengstrand2009, Sobolewska2019}. Surveys have identified sources with dense absorbing mediums \citep{Guainazzi2004,Ostorero2016,Ostorero2017,Sobolewska2019b}, nonthermal emission from the compact jets and lobes \citep{Migliori2016, Principe2020}, and strong thermal emission from the shocked ISM \citep{Siemiginowska2010,Siemiginowska2016}. X-ray properties of the young radio quasar population at high-redshift have yet to be studied in similar depth. Such an analysis would provide a critical look into the evolution of X-ray emission mechanisms from these young radio sources over cosmic time.

Detecting young quasars in radio at high-redshift has historically been challenging given the existing anticorrelation between the radio source age and the peak frequency of its synchrotron spectrum \citep{ODea1998}. The observed peak frequency for quasars at $z > 4.5$ will shift to less than $1\rm\,GHz$, resulting in detection of only the steepest spectra at GHz frequencies. Recent observations with very long baseline interferometry (VLBI) and the Giant Metrewave Radio Telescope have therefore focused on MHz-peaked spectra to detect young radio sources, identifying 15 quasars at $z > 4.5$ with steep radio spectra \citep{Falcke2004,Coppejans2016,Coppejans2017}. These results represent the first high-redshift sample of young radio quasars with known radio morphology and broadband radio spectra, though the X-ray properties of these sources are yet unknown. It is this topic of interest that motivates our study.

This paper presents \chandra\ observations of 15 young radio quasars selected from recent MHz surveys where all targets are at redshift range $4.5 < z < 5.0$, so the X-ray observations are sensitive to $\sim$\,3--40\,keV rest-frame energy range of each target. We investigate for differences, if any, both amongst our sample and in the broader context of previous X-ray surveys. The remainder of the paper is structured as follows. Section~\ref{sect:observation} details the sample selection criteria and the data reprocessing of the observations. Section~\ref{sect:xray} describes the X-ray photometric and spectroscopic analysis of the sample.  Optical and X-ray properties in regards to both the sample and other surveys are provided in Section~\ref{sect:ox}. Radio and X-ray properties of the sample are provided in Section~\ref{sect:radio}, and statistically significant outliers within our sample are discussed in Section~\ref{sect:outlier}. Our concluding remarks are provided in Section~\ref{sect:conclude}. 

For this paper, we adopted $H_{0} = 70\rm\,km\,s^{-1}\,Mpc^{-1}$, $\Omega_{\Lambda} =0.7$, and $\Omega_{M} = 0.3$ \citep{Hinshaw2013}. 

\begin{figure*}
  \begin{tightcenter}   
    \includegraphics[width=0.99\linewidth]{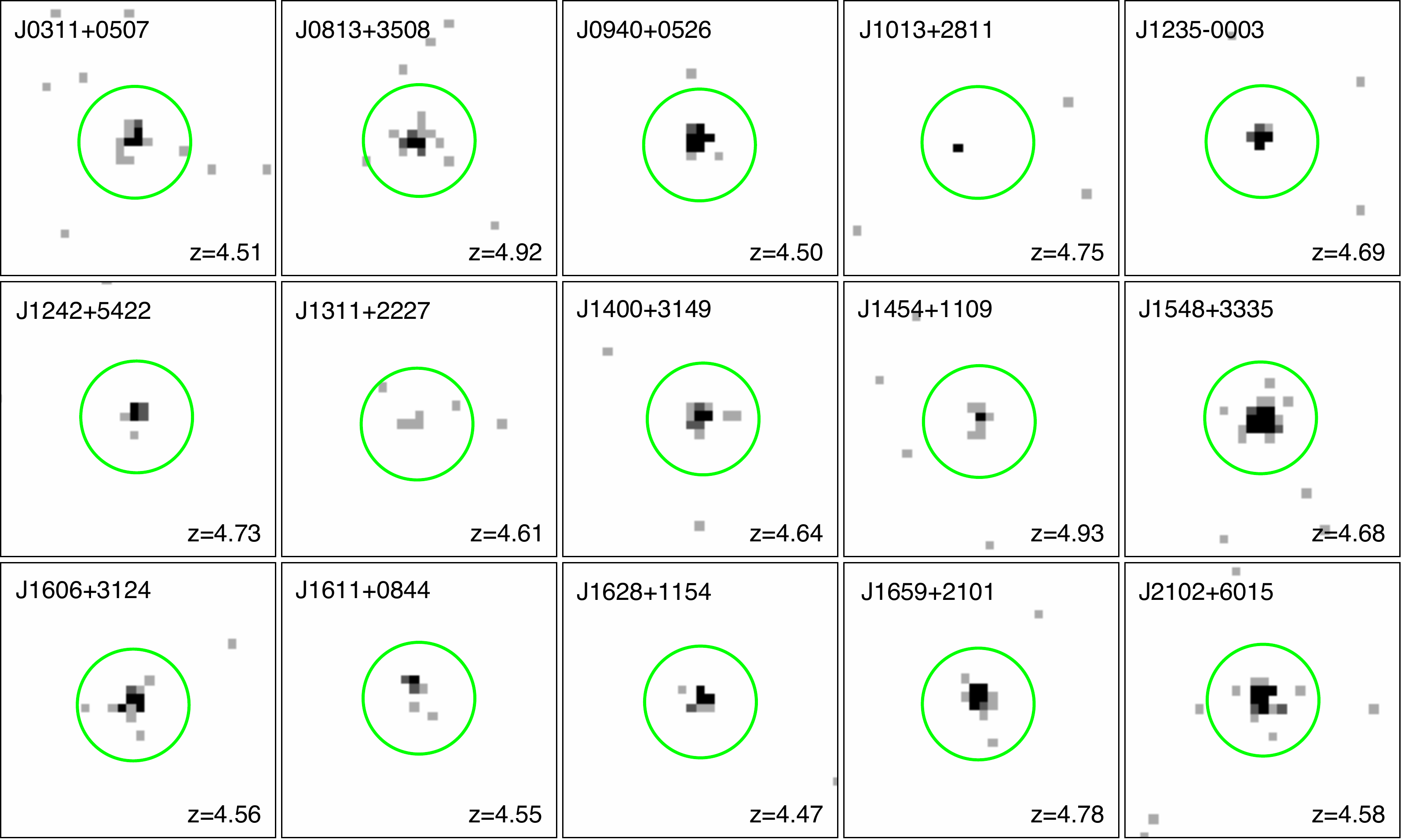}  
  \end{tightcenter}
\caption{0.5--7.0\,keV \chandra{} images of the 15 targets analyzed in this work, where each image is $28\,\times28$ pixels ($13\farcs8\,\times\,13\farcs8$). The 2\farcs95 radius circles (green) were the regions used for the X-ray photometric and spectroscopic analyses discussed in Section~\ref{sect:xray}.}
\label{fig:sample}
\end{figure*}
 
\section{Sample Selection and Data Reduction} 
\label{sect:observation}

Targets for our X-ray survey were chosen from the \cite{Coppejans2016,Coppejans2017} catalog of compact radio quasars at $z > 4.5$ with known spectroscopic redshifts that were previously studied at high angular resolution with VLBI. The catalog totaled 30 unique radio sources, each with broadband radio spectra. We specifically selected sources from the catalog with steep or peaked spectra at MHz frequencies as that indicates relatively young radio sources \citep{ODea1998}. Targets from the radio sample with redshifts between $4.5 < z < 5.0$  were prioritized in order to maximize scientific return with minimal X-ray exposure time. In total, 15 sources from the \cite{Coppejans2016,Coppejans2017} catalog met our selection criteria.
 
A cross-examination of the 15 selected radio sources with archival X-ray databases found only two targets, J0813+3508 and J1242+5422, previously observed in X-rays, both of which were with \chandra{} \citep{Zhu2019}. We therefore selected the remaining 13 targets from the radio sample to observe in X-rays, giving us a total of 15 radio quasars for our analysis. The selected quasar redshift range represents a historically undersampled range in X-ray databases as such targets commonly lie below the sensitivity limits of large sky surveys or snapshot observations \citep[e.g.,][]{Lusso2016}, while deep-exposure X-ray observations generally prioritize the highest-redshift sources at $z>6$ \citep[e.g.,][]{Banados2018,Vito2019}. Thus, a further motivation for our survey is to provide more complete coverage of the redshift parameter space.
 
Each target selected for the X-ray survey was observed with \chandra{} using the Advanced CCD Image Spectrometer (ACIS-S) instrument in {\tt vfaint} mode with the aimpoint on the S3 chip. Young radio quasars generally extend no more than a few kiloparsecs from the nucleus and are therefore predicted to appear unresolved at the 0\farcs5 resolution of \chandra. Nevertheless, an extended emission analysis for each source is described in Section~\ref{sect:extend}. All observations were analyzed using the level 2 data products from the \chandra{} standard data processing pipeline together with the software package \ciao\,v4.11.5 with \caldb\,v4.8.5, where each observation was reprocessed using the routine {\tt deflare} to remove background flaring periods from the data. The average cleaned exposure time per target is 5.95\,ks, and the cleaned exposures were used for the remaining analysis discussed in this article. An overview of the observation information for each target is provided in Table~\ref{table:obs}, and 0.5--7.0\,keV \chandra\ images of the sample are shown in Figure~\ref{fig:sample}.

\section{X-Ray Analysis} 
\label{sect:xray}

\subsection{Hardness Ratios}
\label{sect:hr}

\begin{table*}
	\caption{Observed X-Ray Photometry and Hardness Ratios}
	\label{table:hr}
	\begin{tightcenter}
	\begin{tabular}{ c c c c c  }
		\hline \hline
		Object & \multicolumn{3}{c}{X-Ray Counts} & $HR$\\
		\cline{2-4}
        & 0.5$-$7.0\,keV & 0.5$-$2.0\,keV & 2.0$-$7.0\,keV & \\
		\hline
        J0311$+$0507 & $28.6\substack{+4.7\\-5.7}$ & $17.4\substack{+3.3\\-4.7}$ & $11.2\substack{+2.5\\-4.0}$ & $-0.21\substack{+0.17\\-0.20}$ \\
        J0813+3508 & $29.5\substack{+4.4\\-6.2}$ & $16.3\substack{+3.2\\-4.6}$ & $13.2\substack{+2.9\\-4.2}$ & $-0.10\substack{+0.18\\-0.18}$ \\
        J0940$+$0526 & $35.7\substack{+5.1\\-6.6}$ & $25.4\substack{+4.1\\-5.7}$ & $10.3\substack{+2.5\\-3.8}$ & $-0.42\substack{+0.14\\-0.15}$ \\
        J1013$+$2811 & $3.7\substack{+1.1\\-2.4}$ & $2.4\substack{+0.7\\-1.9}$ & $< 3.6$ & $< +0.36$ \\
        J1235$-$0003 & $16.8\substack{+3.2\\-4.7}$ & $10.4\substack{+2.6\\-3.6}$ & $6.4\substack{+1.8\\-3.0}$ & $-0.24\substack{+0.21\\-0.26}$ \\
        J1242$+$5422 & $13.6\substack{+3.3\\-3.9}$ & $11.4\substack{+2.6\\-3.8}$ & $2.2\substack{+0.6\\-2.0}$ & $-0.67\substack{+0.14\\-0.23}$ \\
        J1311$+$2227 & $6.5\substack{+1.7\\-3.2}$ & $5.3\substack{+1.5\\-2.8}$ & $< 3.6$ & $< -0.18$ \\
		J1400$+$3149 & $19.7\substack{+3.7\\-4.9}$ & $12.4\substack{+2.9\\-3.9}$ & $7.3\substack{+1.9\\-3.3}$ & $-0.26\substack{+0.07\\-0.23}$ \\
        J1454$+$1109	& $9.8\substack{+2.0\\-3.9}$ & $5.4\substack{+1.5\\-2.8}$ & $4.4\substack{+1.2\\-2.6}$ & $-0.12\substack{+0.30\\-0.33}$ \\
        J1548$+$3335 & $78.6\substack{+7.6\\-9.8}$ & $60.3\substack{+7.0\\-8.4}$ & $18.3\substack{+3.6\\-4.8}$ & $-0.53\substack{+0.08\\-0.10}$ \\
        J1606$+$3124 & $30.6\substack{+4.3\\-6.4}$ & $4.4\substack{+1.1\\-2.7}$ & $26.2\substack{+4.1\\-5.6}$ & $+0.71\substack{+0.15\\-0.10}$ \\
        J1611$+$0844 & $10.7\substack{+2.6\\-3.6}$ & $7.4\substack{+2.0\\-3.1}$ & $3.3\substack{+1.0\\-2.3}$ & $-0.39\substack{+0.20\\-0.33}$ \\
        J1628$+$1154 & $19.8\substack{+3.2\\-5.5}$ & $14.4\substack{+3.0\\-4.3}$ & $5.4\substack{+1.5\\-2.8}$ & $-0.45\substack{+0.15\\-0.24}$ \\
        J1659$+$2101 & $33.5\substack{+5.0\\-6.2}$ & $20.3\substack{+3.8\\-5.0}$ & $13.2\substack{+2.9\\-4.2}$ & $-0.21\substack{+0.14\\-0.19}$ \\
        J2102$+$6015 & $48.5\substack{+5.8\\-7.7}$ & $22.4\substack{+4.0\\-5.3}$ & $26.1\substack{+4.4\\-5.7}$ & $+0.08\substack{+0.13\\-0.16}$ \\
		\hline
	\end{tabular}
	\end{tightcenter}
	Background-subtracted counts and hardness ratios ($HR$) for the sources. $HR$ is defined as $\frac{H-S}{H+S}$, where $H$ and $S$ correspond to the hard (2.0--7.0\,keV) and soft (0.5--2.0\,keV) bands, respectively. Errors on X-ray counts and $HR$ were computed to $1\sigma$ according to the method in \cite{Park2006}. Upper bounds were computed to $3\sigma$.
\end{table*}

X-ray emission was measured from each source using a 2\farcs95 radius, circular region centered on the target. The defined source region corresponds to $>$\,$95\%$ the total encircled energy fraction at 1.5\,keV for an on-axis target, and each target analyzed in this work was on-axis ($\theta < 1\arcmin$) in its respective observation. We chose a large source region in order to reliably estimate the count rate, and subsequent flux, without the need of an aperture correction. Given that the contaminant layer on ACIS is known to contribute a systematic count rate $\sim$\,5\%\footnote{See ``ACIS QE Contamination" for further information:\\\url{ https://cxc.harvard.edu/ciao/why/acisqecontamN0010.html}}, this method of count rate measurement is sufficient for our analysis. Background X-ray emission was measured from a circular region that we defined to be adjacent to the target. Each background region was, at minimum, 50 times larger in area than the source region and was verified to be free of X-ray sources. 

In Table~\ref{table:hr}, we present the  background-subtracted counts for each source over the 0.5--2.0\,keV (soft) and 2.0--7.0\,keV (hard) energy bands. The hardness ratio $HR$ was measured for each target using these bands. $HR$ is defined as $\frac{H-S}{H+S}$, where $H$ and $S$ correspond to the background-subtracted counts in the hard and soft bands, respectively. Statistical errors on X-ray counts and $HR$ were computed to $1\sigma$ using a Bayesian estimation of hardness ratios (BEHR), as described in \cite{Park2006}. For sources undetected in the hard band, we provided a $3\sigma$ upper bound on detection as derived from the BEHR analysis. The calculated $HR$ for each source is reported in Table~\ref{table:hr}.

The average $HR$ for the detected sources equals $-0.22\substack{+0.16\\-0.20}$. Examination of the scatter in the $HR$ results shows that object J1606+3124 diverges from the remaining sample by $>3\sigma$ with a $HR$ of $+0.71\substack{+0.15\\-0.10}$. The significant hard X-ray excess over its rest-frame energy range of 3--40\,keV may indicate that J1606 is a highly obscured source. Further discussion of this target is provided in Section~\ref{sect:J1606}. Excluding J1606, the average $HR$ for the remaining detected sources is $-0.29\substack{+0.15\\-0.21}$, and all remaining targets agree with the average value to within $2\sigma$.

\subsection{X-Ray Fluxes and Luminosities}
 \label{sect:lumin}

\begin{figure*}
  \begin{tightcenter}   
    \includegraphics[width=0.44\linewidth]{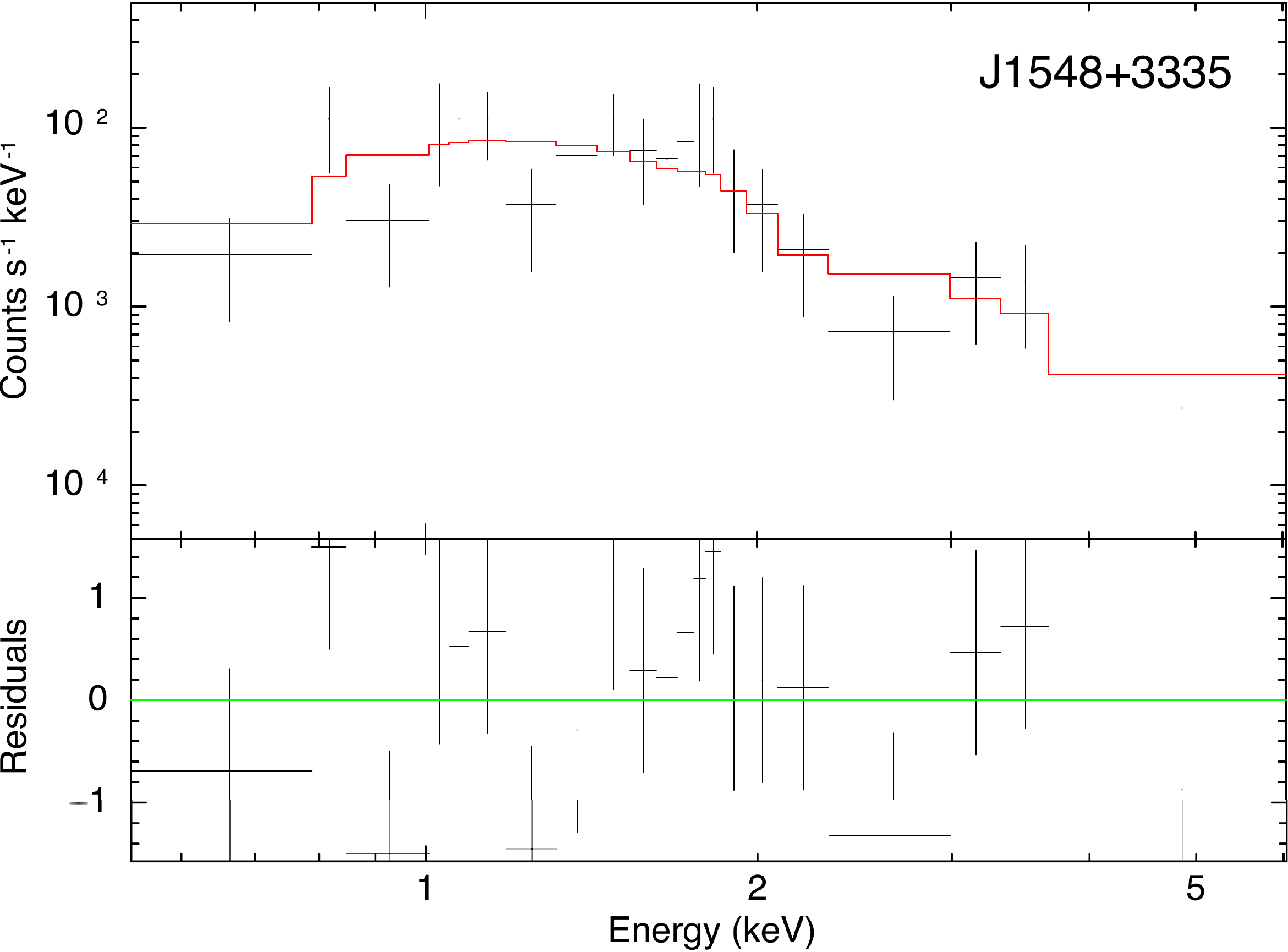} 
    \hspace{3em}
    \includegraphics[width=0.44\linewidth]{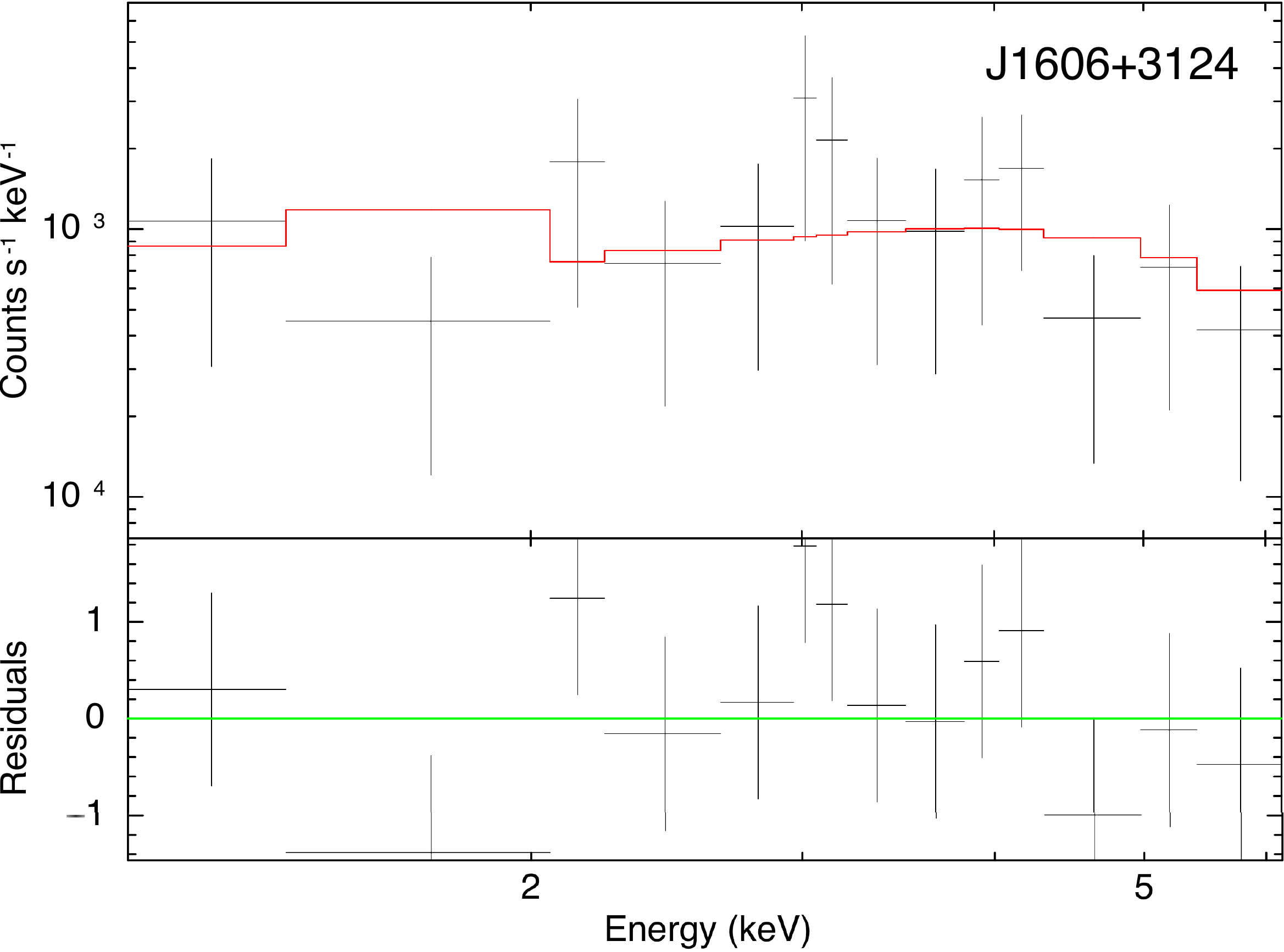} 
  \end{tightcenter}
\caption{\chandra\ X-ray spectra of two sources in the sample, J1548$+$3335 ($\Gamma = 2.17\substack{+0.13\\-0.11}$) and J1606$+$3124 ($\Gamma = -0.11\substack{+0.41\\-0.43}$). The spectra are binned by 1 count per bin and are fitted over the 0.5--7.0\,keV energy range with an absorbed power-law model. The lower panels show the residuals from the best-fit model.}
\label{fig:spectra}
\end{figure*}

To determine unabsorbed fluxes for the sample, we extracted an emission spectrum from each observation using the {\tt specextract} routine in \ciao. Source and background regions were defined to be the same as those in Section~\ref{sect:hr}. Each spectrum was modeled over the 0.5--7.0\,keV band with an absorbed power-law  (i.e., {\tt phabs$\cdot$powerlaw}) using WStat statistics in Sherpa \citep{Freeman2001}. All spectra were binned to have a minimum of 1 count per bin in order to mitigate known issues with background bias in WStat\footnote{See `Bias In Profile Poisson Likelihood,'\\ \url{https://giacomov.github.io/Bias-in-profile-poisson-likelihood/}}. Galactic equivalent hydrogen column densities $N_{\rm H}$ were fixed to values from \cite{Dickey1990}, while the photon index $\Gamma$ ($f_{E} \propto E^{-\Gamma}$) was varied. Spectra for two notable sources, J1548$+$3335 and J1606$+$3124, are shown in Figure~\ref{fig:spectra}. Results from the spectral analysis of the sample, details discussed below, are provided in Table~\ref{table:xray}. 

The Galactic absorption-corrected 0.5--7.0\,keV flux $f_{\rm 0.5\mbox{-}7.0\,keV}$ was measured for each source using the best-fit spectral model. We measured the fluxes for J1548$+$3335 and J1606$+$3224 individually given their unique characteristics (discussed below). Since the remaining 13 sources had low count statistics (designated with parentheses in Table~\ref{table:xray}), we estimated their fluxes using the best-fit photon index $\Gamma$ obtained from a simultaneous fit. The rest-frame 2--10\,keV luminosity $L_{\rm 2\mbox{-}10\,keV}$ and rest-frame, monochromatic 2\,keV luminosity $\ell_{\rm 2\,keV}$ were determined for each source from its best-fit model, and the results are shown in Table~\ref{table:xray}.

{\bf J1548$+$3335} was modeled individually given its high count statistics when compared to the remainder of the sample, 79 counts over the 0.5--7.0\,keV energy range. The best-fit photon index was $\Gamma = 2.17\substack{+0.13\\-0.11}$ with WStat fit statistics of 45.9 for 63 degrees of freedom (dof). An additional intrinsic absorption component (i.e., {\tt zphabs}) was added to the model, and no appreciable intrinsic absorption $N_{\rm H}^i$ was measured. A 3$\sigma$ upper bound for $N_{\rm H}^i$ was determined to be $4.1 \times 10^{23}\rm\,cm^{-2}$. 

{\bf J1606$+$3224} spectrum was also modeled individually as it was confirmed to be a statistically significant outlier from the $HR$ analysis.  Based on the significant soft X-ray extinction observed from its $HR$, we fit J1606 with a {\tt phabs$\cdot$(zphabs$\cdot$powerlaw}) expression. Initial attempts to simultaneously fit the $\Gamma$ and $N_{\rm H}^i$ parameters failed due to the low count statistics for the observed spectrum. $N_{\rm H}^i$ was subsequently fixed at zero, producing a best-fit of $\Gamma = -0.11\substack{+0.41\\-0.43}$ with WStat fit statistics/dof = 20.5/26. Alternatively, the spectral analysis was repeated with the photon index fixed to the best-fit value from the remaining sample, $\Gamma = 1.52$, derived below, and $N_{\rm H}^i$ free to vary. With this model, the best-fit intrinsic column density is $1.34\substack{+0.45\\-0.38} \times 10^{24}\rm\,cm^{-2}$ with WStat fit statistics/dof = 19.5/26. The statistics from the two models are equivalent to one another, so neither model is favored. The unabsorbed fluxes were consistent between the models, so we utilized the model with no intrinsic absorption for the remainder of the X-ray analysis. A discussion of the different spectral models and their implications is provided in Section~\ref{sect:J1606}. 

{\bf The remaining sources} were initially modeled separately, and the best-fit $\Gamma$ for each source is listed in Table~\ref{table:xray}. However, the individual fits placed poor constraints on the model parameters. The 13 remaining sources were therefore simultaneously fit where both $\Gamma$ and $N_{\rm H}^i$ were tied between the spectra while the normalization was allowed to vary, which gave a best-fit of $\Gamma = 1.52\substack{+0.13\\-0.14}$ with WStat fit statistics/dof = 184.4/214. No intrinsic absorption was detected from any of the 13 sources, and a 3$\sigma$ upper bound for $N_{\rm H}^i$ from the simultaneous best-fit was measured as $1.1 \times 10^{23}\rm\,cm^{-2}$. 

\begin{table*}
	\caption{Optical and X-ray Properties of the Sample}
	\label{table:xray}
	\begin{tightcenter}
	\begin{tabular}{ c c c c c c c c c }
		\hline \hline
		Object & $N_{\rm H}$ & $\Gamma$ & $f_{\rm 0.5\mbox{-}7.0\,keV}$ & $L_{\rm 2\mbox{-}10\,keV}$ & $\ell_{\rm 2\,keV}$ & $m_{1450\rm\,\angstrom}$ & $\ell_{2500\rm\,\angstrom}$ & $\alpha_{\rm ox}$ \\
		(1) & (2) & (3) & (4) & (5) & (6)  & (7) & (8) & (9) \\
		\hline
        J0311$+$0507& 10.86 & ($0.92\substack{+0.50\\-0.51}$)
        & $6.6\substack{+1.3\\-1.2}$ & $5.3\substack{+2.0\\-1.5}$ & $4.5\substack{+2.6\\-1.7}$ & 22.3\tablenotemark{a} & 1.18 & $-1.31\substack{+0.08\\-0.08}$ \\
        J0813$+$3508 & 4.91 & 
        ($1.33\substack{+0.42\\-0.42}$) & $5.0\substack{+1.0\\-0.8}$ & $4.8\substack{+1.9\\-1.4}$ & $4.1\substack{+2.4\\-1.6}$ & 19.04 & 29.06 & $-1.86\substack{+0.08\\-0.08}$ \\
        J0940$+$0526 & 3.56 & ($2.00\substack{+0.35\\-0.35}$) & $7.5\substack{+1.4\\-1.2}$ & $6.0\substack{+2.1\\-1.6}$ & $5.1\substack{+2.7\\-1.8}$ & 20.61 & 5.54 & $-1.55\substack{+0.08\\-0.08}$ \\
        J1013$+$2811 & 2.56 & (---) & $0.6\substack{+0.5\\-0.3}$ & $0.5\substack{+0.6\\-0.3}$ & $0.4\substack{+0.6\\-0.3}$ & 21.25 & 3.50 & $-1.88\substack{+0.14\\-0.17}$\\
        J1235$-$0003 & 1.90 & ($0.99\substack{+0.63\\-0.63}$) & $4.0\substack{+1.1\\-0.9}$ & $3.5\substack{+1.6\\-1.2}$ & $3.0\substack{+2.0\\-1.2}$ & 20.06 & 10.12 & $-1.74\substack{+0.09\\-0.09}$ \\
        J1242$+$5422 & 1.55 & ($3.59\substack{+1.24\\-1.15}$) & $3.4\substack{+1.0\\-0.8}$ & $3.0\substack{+1.5\\-1.0}$ & $2.5\substack{+1.7\\-1.1}$ & 19.89 & 12.18 & $-1.80\substack{+0.09\\-0.10}$ \\        
        J1311$+$2227 & 1.67 & ($3.09\substack{+1.71\\-1.37}$) & $1.2\substack{+0.6\\-0.5}$ & $1.0\substack{+0.7\\-0.5}$ & $0.9\substack{+0.8\\-0.5}$ & 20.28 & 7.93 & $-1.91\substack{+0.11\\-0.13}$ \\
		J1400$+$3149 & 1.25 & ($1.39\substack{+0.55\\-0.56}$) & $4.0\substack{+1.0\\-0.9}$ & $3.4\substack{+1.5\\-1.1}$ & $2.9\substack{+1.9\\-1.2}$ & 20.17 & 8.91 & $-1.72\substack{+0.08\\-0.09}$ \\
        J1454$+$1109	& 2.00 & ($2.04\substack{+0.97\\-0.91}$) & $1.9\substack{+0.7\\-0.6}$ & $1.8\substack{+1.1\\-0.7}$ & $1.5\substack{+1.2\\-0.7}$ & 21.02 & 4.67 & $-1.72\substack{+0.10\\-0.11}$ \\
        J1548$+$3335 & 2.25 & $2.17\substack{+0.13\\-0.11}$ & $14.8\substack{+1.7\\-1.7}$ & $23.2\substack{+7.5\\-6.1}$ & $34.0\substack{+19.1\\-13.0}$ & 20.65 & 5.82 & $-1.24\substack{+0.07\\-0.08}$\\
        J1606$+$3124 & 2.59 & $-0.11\substack{+0.41\\-0.43}$ & $9.8\substack{+2.0\\-1.7}$ & $1.1\substack{+1.5\\-0.5}$ & $0.2\substack{+0.6\\-0.1}$ & 20.89 & 4.40 & $-2.08\substack{+0.24\\-0.16}$ \\
        J1611$+$0844 & 4.06 & ($2.46\substack{+1.41\\-1.02}$) & $2.1\substack{+0.8\\-0.6}$ & $1.7\substack{+1.0\\-0.7}$ & $1.5\substack{+1.1\\-0.7}$ & 19.09 & 22.84 & $-1.99\substack{+0.10\\-0.10}$ \\
        J1628$+$1154 & 5.12 & ($1.79\substack{+0.58\\-0.56}$) & $4.7\substack{+1.2\\-1.0}$ & $3.7\substack{+1.6\\-1.2}$ & $3.1\substack{+2.0\\-1.3}$ & 21.01 & 3.76 & $-1.57\substack{+0.08\\-0.09}$ \\
        J1659$+$2101 & 5.47 & ($1.41\substack{+0.40\\-0.40}$) & $7.0\substack{+1.3\\-1.1}$ & $6.3\substack{+2.3\\-1.8}$ & $5.4\substack{+3.0\\-2.0}$ & 19.98 & 11.33 & $-1.66\substack{+0.07\\-0.08}$ \\	
        J2102$+$6015 & 39.86 & ($1.27\substack{+0.29\\-0.29}$) & $13.4\substack{+2.1\\-1.8}$ & $11.1\substack{+3.7\\-2.8}$ & $9.4\substack{+4.8\\-3.3}$ & 21.44\tablenotemark{b} & 2.69 & $-1.33\substack{+0.07\\-0.07}$\\
		\hline
	\end{tabular}
	\end{tightcenter}
	(1) Object name. (2) Galactic column density extrapolated from \cite{Dickey1990} in units of $10^{20}\rm\,cm^{-2}$. (3) Photon index estimated from 0.5--7.0\,keV spectral best fit. Values in parentheses correspond to observations with low count statistics, so the average photon index $\Gamma = 1.52\substack{+0.13\\-0.14}$ was used for the remaining analysis. (4) Observed 0.5--7.0\,keV flux in units of $10^{-14}\rm\,erg\,cm^{-2}\,s^{-1}$. (5) Rest-frame 2--10\,keV luminosity in units of $10^{45}\rm\,erg\,s^{-1}$. (6) Rest-frame, monochromatic luminosity at 2\,keV in units of $10^{27}\rm\,erg\,s^{-1}\,Hz^{-1}$. (7) Rest-frame, monochromatic apparent AB magnitude at 1450\,\angstrom, as measured from the $z$-filter of Pan-STARRS. (8) Rest-frame, monochromatic luminosity at 2500\,\angstrom\ in units of $10^{31}\rm\,erg\,s^{-1}\,Hz^{-1}$. (9) Optical--X-ray power-law slope.
    \tablenotemark{a}{Measurement from \cite{Kopylov2006}.}
    \tablenotemark{b}{Extrapolated from ZTF $r$-band photometry.}
\end{table*}

\subsection{Extended X-Ray Emission}
\label{sect:extend}

We additionally studied each target within the sample for evidence of extended X-ray emission. Ray-tracing files were simulated for each observation using ChaRT v2 with MARX v5.4.0, and a synthetic point-source function (PSF) image was generated with the {\tt simulate\_psf} routine in \ciao. A surface brightness profile was measured for each object out to a radius of 4\farcs92, where the defined annular sectors were centered on the AGN. The sectors were divided into $10^{\circ}$ segments, and each annular region had a radial width of 0\farcs492 to match the pixel size of \chandra\ ACIS. Radial profiles of the 0.5--7.0\,keV observations were compared against the simulated PSF images for evidence of elevated asymmetric count rates from each source.

We initially detected extended emission to the east of J1606$+$3124, but the feature was determined to be coincident in size and position with a known \chandra\ PSF artifact\footnote{See the PSF artifact caveat in the \ciao{} User Guide \\ \url{https://cxc.harvard.edu/ciao/caveats/psf_artifact.html}} and was therefore considered nonphysical. No evidence of extended emission was detected from any other source above the average $3\sigma$ background limit of $4\times10^{-15}\rm\,erg\,cm^{-2}\,s^{-1}$ over the observed 0.5--7.0\,keV energy range. Thus, we concluded that no detectable extension was found in any target within the sample.

\section{Optical and X-Ray Properties} 
\label{sect:ox}

\begin{figure*}
    \begin{tightcenter}
    \includegraphics[width=0.497\textwidth]{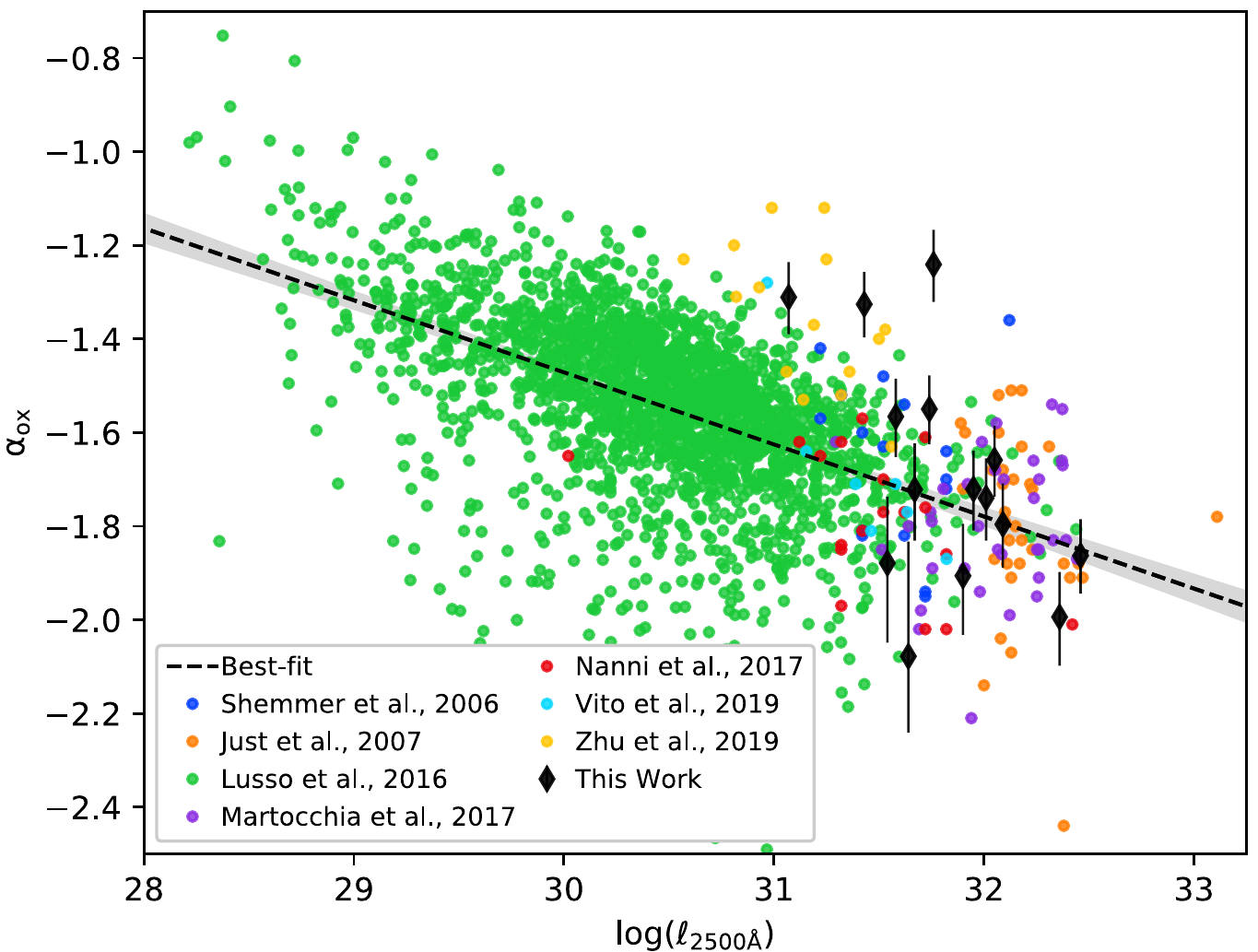} 
    \includegraphics[width=0.497\textwidth]{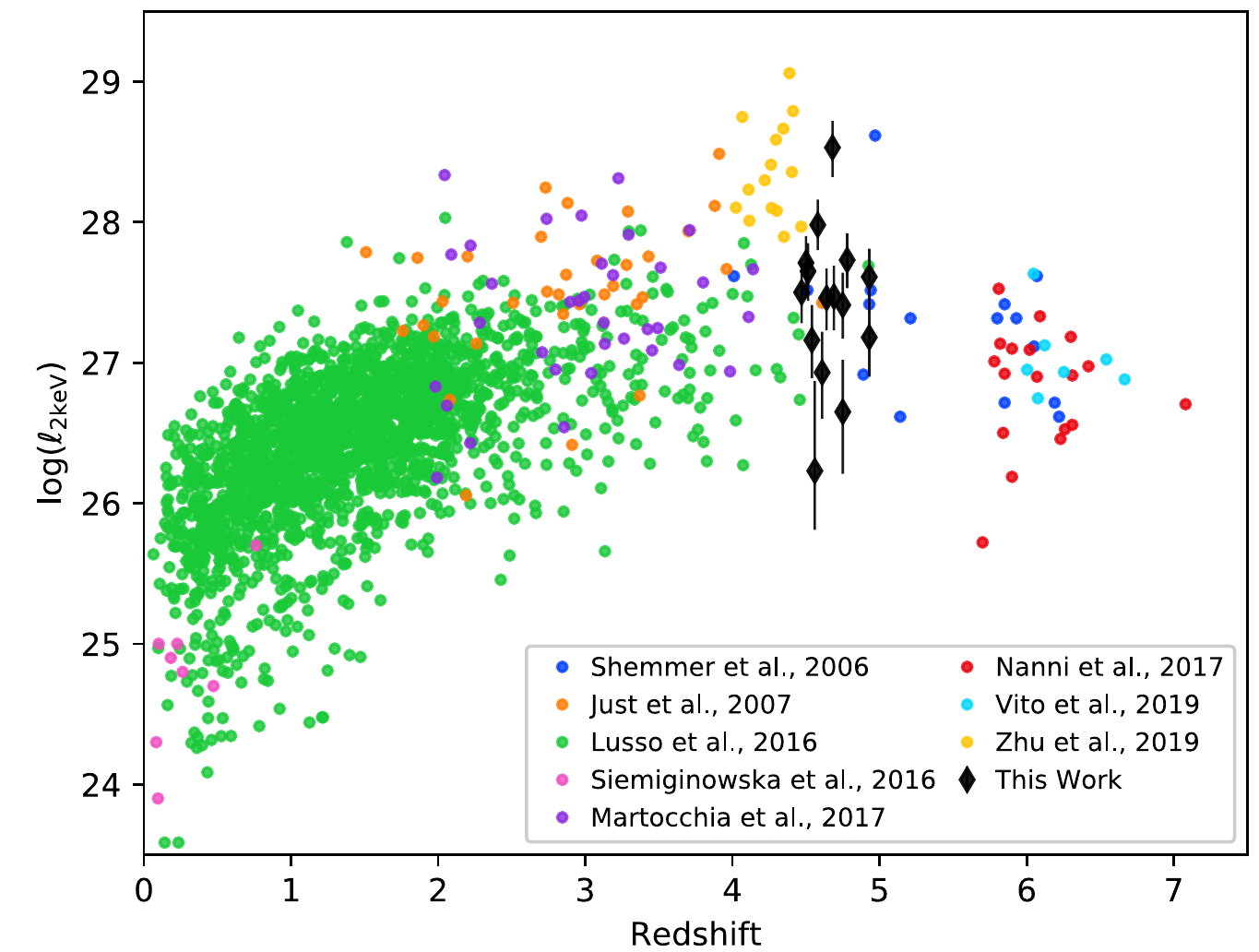}
    \includegraphics[width=0.497\textwidth]{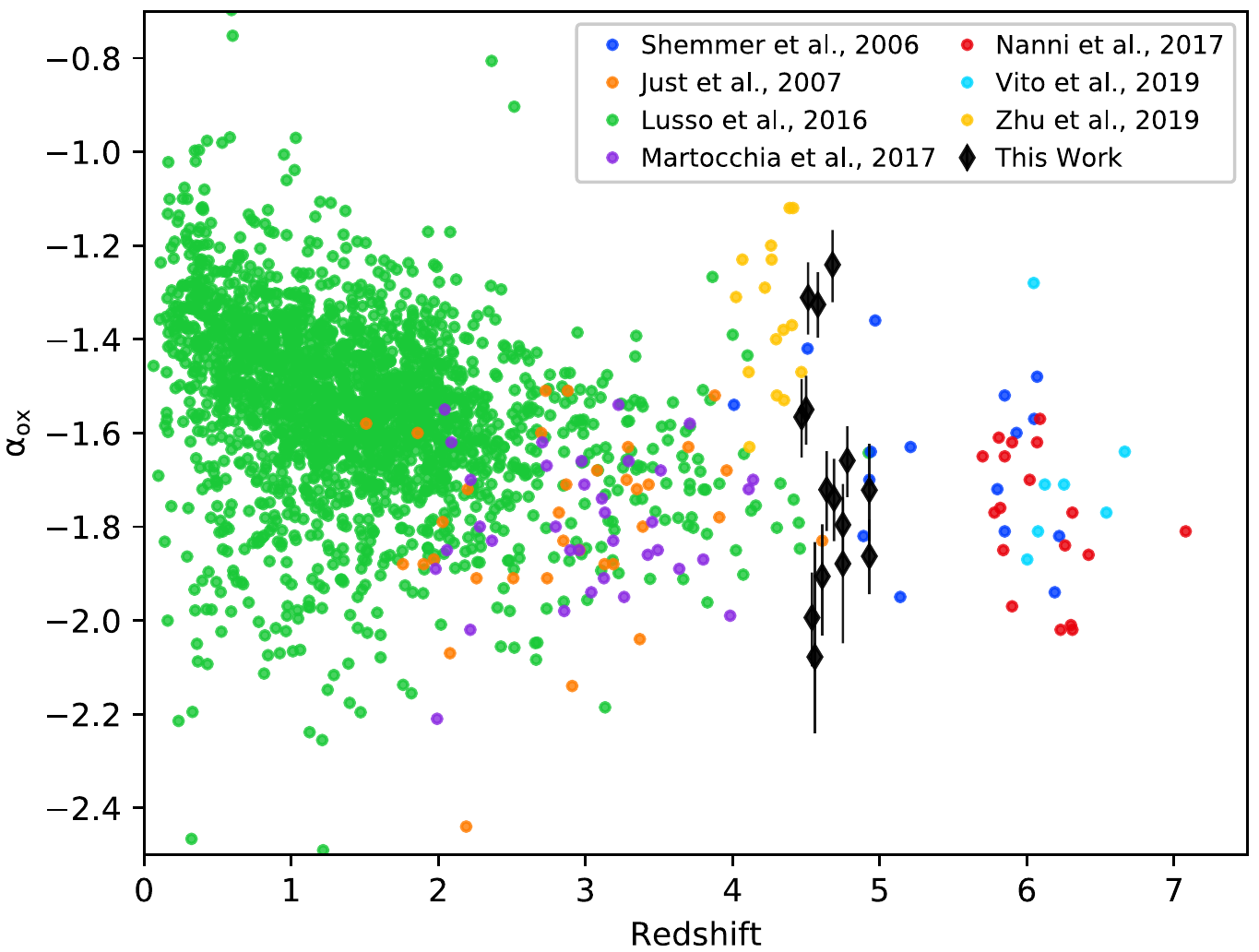}
    \includegraphics[width=0.497\textwidth]{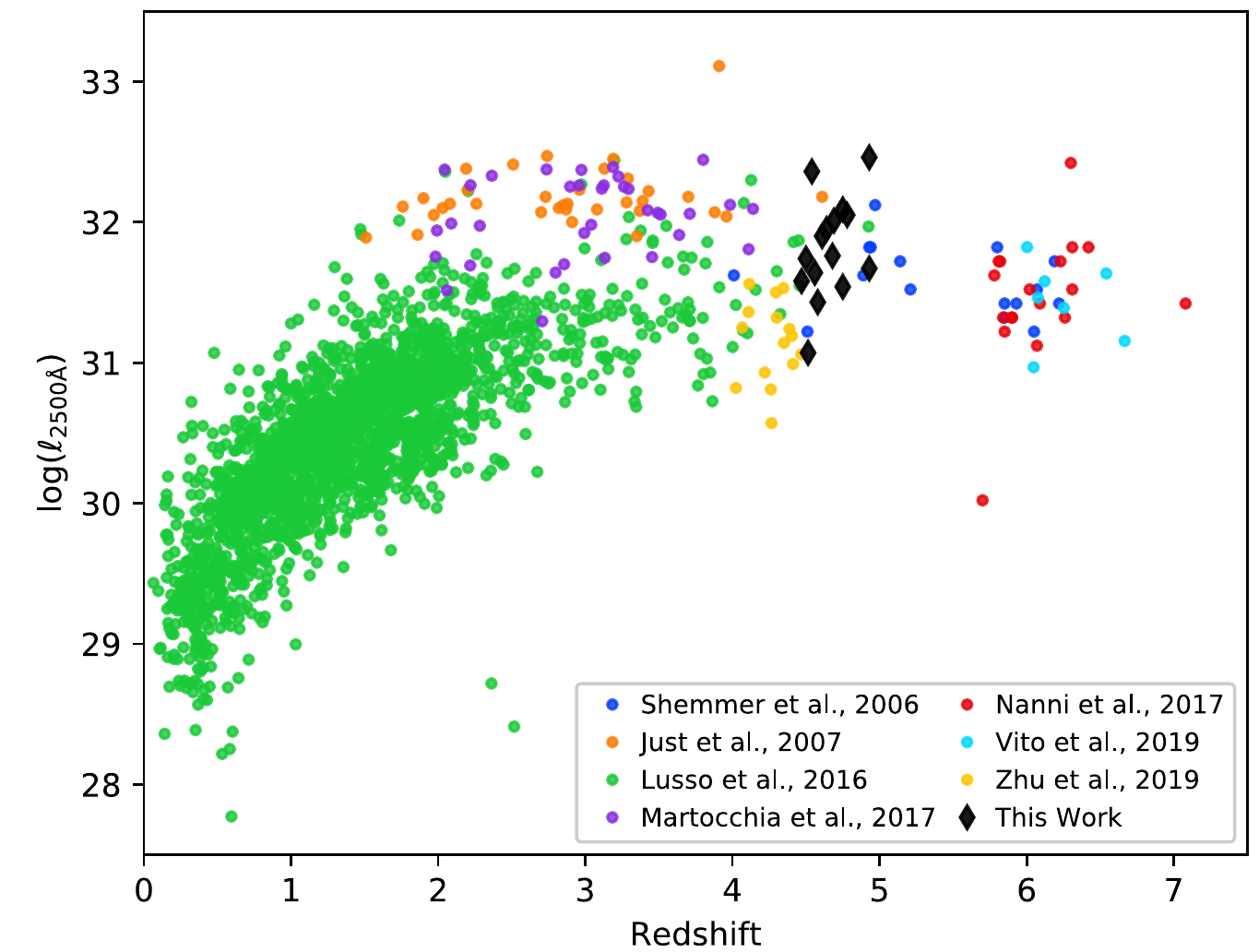}
    \end{tightcenter}
\caption{{\bf Top Left}: optical--X-ray power-law slope $\alphaox$ versus UV luminosity $\ell_{2500\rm\,\angstrom}$. {\bf Top Right}: X-ray luminosity $\ell_{2\rm\,keV}$ versus redshift. {\bf  Bottom Left}: $\alphaox$ versus redshift. {\bf Bottom Right}: $\ell_{2500\rm\,\angstrom}$ versus redshift. Results from this work (black) are compared against other quasar samples from the literature. The black dotted line is the overall best-fit, while the grey region is the $3\sigma$ confidence level. Previous measurements are taken from \cite{Shemmer2006,Just2007,Lusso2016,Siemiginowska2016,Nanni2017,Martocchia2017,Zhu2019,Vito2019b}.}
\label{fig:alpha}
\end{figure*}

Historically, quasar studies have found the optical--X-ray flux ratio to be inversely dependent on the optical luminosity as well as independent of redshift \citep[e.g.,][]{Avni1982,Tananbaum1986,Wilkes1994}. We therefore investigated the optical--X-ray relationship for our sample by compiling the optical and UV luminosities from the literature.

In keeping with independent studies of high-redshift AGN, we focused our analysis on the rest-frame $1450\,\angstrom$ and $2500\,\angstrom$ continuum luminosities. For AGN at $4.5 < z < 5.0$, the rest-frame emission at $1450\,\angstrom$ corresponds to an observed wavelength range of 7975--8700$\,\angstrom$. Therefore, we used the monochromatic AB apparent magnitude from the Pan-STARRS $z$-band for $m_{1450\rm\,\angstrom}$ measurements as the mean wavelength of the $z$-filter corresponds to $8690.1\,\angstrom$ \citep{Tonry2012}. Pan-STARRS observations provided apparent magnitudes for 13 sources from our sample. No Galactic extinction correction was applied to the magnitude measurements. Magnitude measurements of J0311$+$0507 were obtained from \cite{Kopylov2006}. A literature search for J2102$+$6015 provided only an $r$-band photometric measurement with the Zwicky Transient Facility \citep[ZTF;][]{Masci2018}. Since $z$-band photometry was required for our analysis, we extrapolated the ZTF measurement using the average $r$--$z$ color from Pan-STARRS for the remainder of the sample, which was determined to be 1.05. Thus, we obtained photometric optical information for all sources in our sample. 

Direct measurements of the rest-frame $2500\,\angstrom$ luminosity were unavailable in the literature as the observed wavelengths are outside standard survey bands. Thus, we extrapolated the monochromatic luminosity $\ell_{2500\rm\,\angstrom}$ for each object from its apparent magnitude $m_{1450\rm\,\angstrom}$. A UV spectral index $\alpha=-0.5$ was assumed ($f_{\nu} \propto \nu^{\alpha}$), which is consistent with prior studies \citep[i.e.,][]{Shemmer2006,Just2007,Nanni2017}.  Results for $m_{1450\rm\,\angstrom}$ and $\ell_{\rm 2500\angstrom}$ are provided in Table~\ref{table:xray}.

Having determined both the optical and X-ray luminosities for the sample, we calculated the optical--X-ray power-law index $\alphaox$ for each source. In keeping with \cite{Tananbaum1979}, we defined $\alphaox$ as
\begin{equation}
\frac{{\rm log}(\ell_{\rm 2\,keV}/\ell_{\rm 2500\,\angstrom})}{{\rm log}(\nu_{\rm 2\,keV}/\nu_{\rm 2500\,\angstrom})} = 0.3838\cdot\rm log(\ell_{2\,keV}/\ell_{\rm 2500\,\angstrom}),
\end{equation}
where $\ell_{\rm 2\,keV}$ and $\ell_{\rm 2500\angstrom}$ correspond to the monochromatic luminosities at 2\,keV and 2500\,\angstrom, respectively. To calculate the errors on $\alphaox$, the X-ray count uncertainty was summed in quadrature with an approximated 10\% optical uncertainty, or $\sim$\,0.1 mag. Results for $\alphaox$ are provided in Table~\ref{sect:xray}.

\subsection{Comparison of Optical Properties to Other Surveys}
\label{sect:aox_compare}

Previous analyses of quasars demonstrated an anticorrelation between $\alphaox$ and $\ell_{2500\angstrom}$ \citep{Bechtold1994,Vignali2003, Steffen2006, Kelly2007, Nanni2017}. To determine the impact of our radio-loud quasar sample on this known relationship, we compared our $\alphaox$\,--\,$\ell_{2500\angstrom}$ values against results from other X-ray bright quasar surveys. Given the limited amount of additional high-redshift, radio-loud quasars with known X-ray properties (see Section~\ref{sect:radioloud}), we selected both radio-loud and radio-quiet quasars from the literature. We included 16 targets from \cite{Shemmer2006}, 34 from \cite{Just2007}, 2153 from \cite{Lusso2016}, 18 from \cite{Nanni2017}, 35 from \cite{Martocchia2017}, 15 from \cite{Zhu2019}, and 7 from \cite{Vito2019b}. Two sources were omitted from the \cite{Zhu2019} sample as they are also present in our analysis, though we note that the different analyses yield consistent results in both X-ray and optical properties. Together with the sources from our analysis, we obtained a final sample of 2293 quasars with known $\alphaox$ and $\ell_{2500\angstrom}$ parameters.

The $\alphaox$\,--\,$\ell_{2500\angstrom}$ relationship is shown in the top left panel of Figure~\ref{fig:alpha}, where an anticorrelation trend is clearly observed. The $\alphaox$ values for our radio-selected sample appear consistent with the overall trend, and their scatter is also consistent with other surveys. We performed a linear regression on the total dataset using the {\tt scipy} python package \citep{Virtanen2020}, giving a best-fit relation between $\alphaox$ and $\ell_{2500\angstrom}$ of
\begin{equation}
    \alphaox = (-0.154\pm0.005)\,{\rm log}(\ell_{2500\angstrom}) + (3.2\pm0.2),
\end{equation}
where the reported errors are 1$\sigma$. The best-fit parameters are consistent with results from \cite{Nanni2017} within  $1\sigma$, despite our different sampling rates of the redshift parameter space. We repeated the fit with just the quasar sample from this work, finding a slope of $-0.2\pm3.1$ and an intercept of $0.05\pm 0.14$, where the large errors are primarily due to our small sample size. As a result, our radio-loud sample agrees with the total trend within $1\sigma$.

\subsection{Redshift Trends in Optical and X-ray Data}
\label{sect:lum_compare}

After compiling X-ray and optical data for quasars over a large redshift range, we investigated the presence, if any, of a redshift dependence. Monochromatic luminosities $\ell_{\rm 2\,keV}$ and $\ell_{2500\angstrom}$ are each compared against redshift in Figure~\ref{fig:alpha}. We included nine low-redshift, compact radio sources from \cite{Siemiginowska2016} in order to track the X-ray properties of quasars down to $z \simeq 0.1$. 

Upon comparing the various surveys, we found that the sample from this work exhibits an above-average optical emission relative to other sources at comparable redshifts. However, we note that this difference may be attributable to the extrapolation technique utilized for the optical measurements in this work (see Section~\ref{sect:ox}). The average X-ray luminosity $\ell_{\rm 2\,keV}$ from the sample is consistent with other sources at $z > 2$, and the overall scatter in both X-ray and optical luminosities from our sample agree with the observed distributions from other surveys. Altogether, the luminosity properties for our quasar sample are broadly consistent with previous quasar studies. 

In addition to studying individual X-ray and optical trends, we checked for a redshift dependence with the X-ray--optical power-law slope $\alphaox$. A comparison between redshift and  $\alphaox$ is shown in the bottom left panel of Figure~\ref{fig:alpha}, and no clear trend is detected between the two parameters. The best-fit relation derived in Section~\ref{sect:aox_compare} was also verified to be independent of redshift for the tested range of $z < 7$. These findings further reinforce previous reports that the optical--X-ray power-law is independent of redshift \citep[e.g.,][]{Lusso2016,Nanni2017,Vito2019b}.

Further inspection of the data show a large gap in known X-ray bright quasars over the $5.00 < z < 5.75$ range. Proper sampling of the redshift parameter space is imperative for an accurate study of AGN evolution and feedback processes. Given the success of the work discussed in this paper to detect high-redshift X-ray sources with minimal exposure times, future X-ray surveys targeting this range may also benefit from utilizing radio studies of MHz-peaked sources when selecting target candidates. 

\section{Radio Properties}
\label{sect:radio}

Radio-loud quasars, on average, are more X-ray luminous than radio-quiet quasars for a given optical luminosity \citep[e.g.,][]{Worrall1987,Wu2013}. Thus, we investigated the relation between the radio and X-ray luminosities for our radio-loud sample. The rest-frame 5\,GHz flux densities were estimated from the radio spectra best fits in \cite{Coppejans2017}. For fainter sources in the sample where the radio spectra were not well-defined ($<$\,10\,mJy at 1.4\,GHz), the observed 1.4\,GHz flux densities were adopted (i.e., $\alpha = 0$ is assumed between rest-frame frequencies of $\sim$\,5--8\,GHz). Given the steep and/or peaked spectra of these sources, our approximation should be treated as an upper limit on the rest-frame 5\,GHz flux for the dim quasars. 

Radio loudness was calculated the same as \cite{Kellerman1989}: $\rl = f_{\rm5\,GHz}/f_{\rm4400\,\angstrom}$, where $f_{\rm5\,GHz}$ and $f_{\rm4400\,\angstrom}$ correspond to the rest-frame flux density at 5\,GHz and 4400\,\angstrom, respectively. Rest-frame 4400\,\angstrom\ flux density was extrapolated from the $m_{1450\rm\,\angstrom}$ measurements assuming a spectral index of $\alpha=-0.5$. As the extrapolated optical values do not account for sources of photoelectric absorption, the $\rl$ estimates should be regarded as upper limits. Radio properties for the sample are provided in Table~\ref{table:radio}. We note that the reported radio luminosities represent the total emission for each source. 

The radio and X-ray luminosities for the sample are shown in Figure~\ref{fig:radio}. We performed a linear regression on the dataset using the method described in \cite{Kelly2007b}, and the best-fit relation was
\begin{equation}
{\rm log}(L_{\rm5\,GHz}) = (0.9\pm0.6){\rm log}(L_{\rm2-10\,keV}) + (4\pm26),
\end{equation}
where the reported errors are $1\sigma$. The measured correlation between the radio and X-ray luminosities is consistent with estimates for young radio quasars at $z < 2$ \citep{Fan2016}, which suggests that the radio and X-ray luminosity correlation is redshift-independent. This result further supports the lack of spectral evolution for quasars over a broad redshift range. Figure~\ref{fig:radio} additionally shows that three sources (J0311+0507, J1548+3335, and J1606+3124) are located away the from the best-fit in excess of $1\sigma$. A discussion of each outlier is provided in Section~\ref{sect:outlier}. 

\subsection{Radio-Loudness and Photon Index}
\label{sect:radioloud}

\begin{table}
	\caption{Radio Properties of the Sample}
	\label{table:radio}
	\begin{tightcenter}
	\begin{tabular}{ c c c c c }
		\hline \hline
		Object & $f_{\rm5\,GHz}$ & $\ell_{\rm5\,GHz}$ & log($\mathcal{R}$) & log($\ell_{\rm5\,GHz}$/$\ell_{\rm2\,keV}$)\\
		(1) & (2) & (3) & (4) & (5) \\
		\hline
        J0311$+$0507 & 915 & 149.1 & 5.1 & $7.5\pm0.3$ \\
        J0813$+$3508 & 58 & 11.6 & 2.6 & $6.5\pm0.3$ \\
        J0940$+$0526 & 80 & 12.9 & 3.3 & $6.4\pm0.2$ \\
        J1013$+$2811 & $14^{*}$ & 2.6 & 2.8 & $6.8\pm0.4$ \\
        J1235$-$0003 & $18^{*}$ & 3.2 & 2.5 & $6.0\pm0.3$ \\
        J1242$+$5422 & 25 & 4.6 & 2.6 & $6.3\pm0.3$ \\
        J1311$+$2227 & $8^{*}$ & 1.4 & 2.2 & $6.2\pm0.3$ \\
		J1400$+$3149 & 25 & 4.4 & 2.7 & $6.2\pm0.2$ \\
        J1454$+$1109 & $10^{*}$ & 2.0 & 2.6 & $6.1\pm0.3$ \\
        J1548$+$3335 & 50 & 8.9 & 3.2 & $5.4\pm0.3$ \\
        J1606$+$3124 & 466 & 77.8 & 4.2 & $8.7\pm0.4$ \\
        J1611$+$0844 & $9^{*}$ & 1.5 & 1.8 & $6.0\pm0.3$ \\
        J1628$+$1154 & 62 & 9.9 & 3.4 & $6.5\pm0.3$ \\
        J1659$+$2101 & 45 & 8.4 & 2.8 & $6.2\pm0.2$ \\
        J2102$+$6015 & 326 & 55.0 & 4.3 & $6.8\pm0.3$ \\
		\hline
	\end{tabular}
	\end{tightcenter}
	(1) Object name. (2) Rest-frame 5\,GHz flux density estimated from spectral best fits in \cite{Coppejans2017} in units of $\rm mJy$. In cases where no best fit was found (denoted with a ${}^{*}$), the observed 1.4\,GHz flux densities were used. (3) Monochromatic luminosity at 5\,GHz in units of $10^{33}\rm\,erg\,s^{-1}\,Hz^{-1}$.  (4) Radio loudness $\mathcal{R}$, defined as $f_{\rm5\,GHz}/f_{\rm4400\,\angstrom}$. (5) Rest-frame radio--X-ray ratio. Flux densities and luminosities are assumed to have a 15\% uncertainty. Radio--X-ray ratio errors include both radio and X-ray uncertainties. 
\end{table}

X-ray spectra of radio-loud quasars, on average, have lower photon indices than radio-quiet quasars \citep{Wu2013}. This may indicate the presence of an additional jet contribution, denser intrinsic column densities, or a combination thereof. We sought to verify this photon index relationship with our radio-loud sample by comparing its best-fit photon index of $\Gamma = 1.52\substack{+0.13\\-0.14}$ to spectral indices from other quasar surveys. 

To ensure our comparison is between equivalent targets, we used surveys of X-ray-bright quasars at $z>4$: three radio-quiet samples \citep{Martocchia2017, Nanni2017, Vito2019b}, and one radio-loud sample \citep{Zhu2019}. \cite{Nanni2017} found that radio-quiet sources have an average photon index of $\Gamma = 1.93\substack{+0.30\\-0.29}$, and measurements from \cite{Martocchia2017} and \cite{Vito2019b} agree within $1\sigma$. This value is greater than the best-fit $\Gamma$ from our sample by $1\sigma$. In addition, the average photon index from the radio-loud sample in \cite{Zhu2019} is $\Gamma = 1.44\substack{+0.47\\-0.18}$, which is consistent with the best-fit $\Gamma$ from this work to within $1\sigma$. The X-ray photon indices of our radio-loud quasars are therefore consistent with another radio-loud sample, and our values are lower than the average photon index of radio-quiet sources at a marginal significance.

\subsection{Comparison with Low-Redshift Sources}
\label{sect:lowz}

\begin{figure}
  \begin{tightcenter}   
    \includegraphics[width=0.99\linewidth]{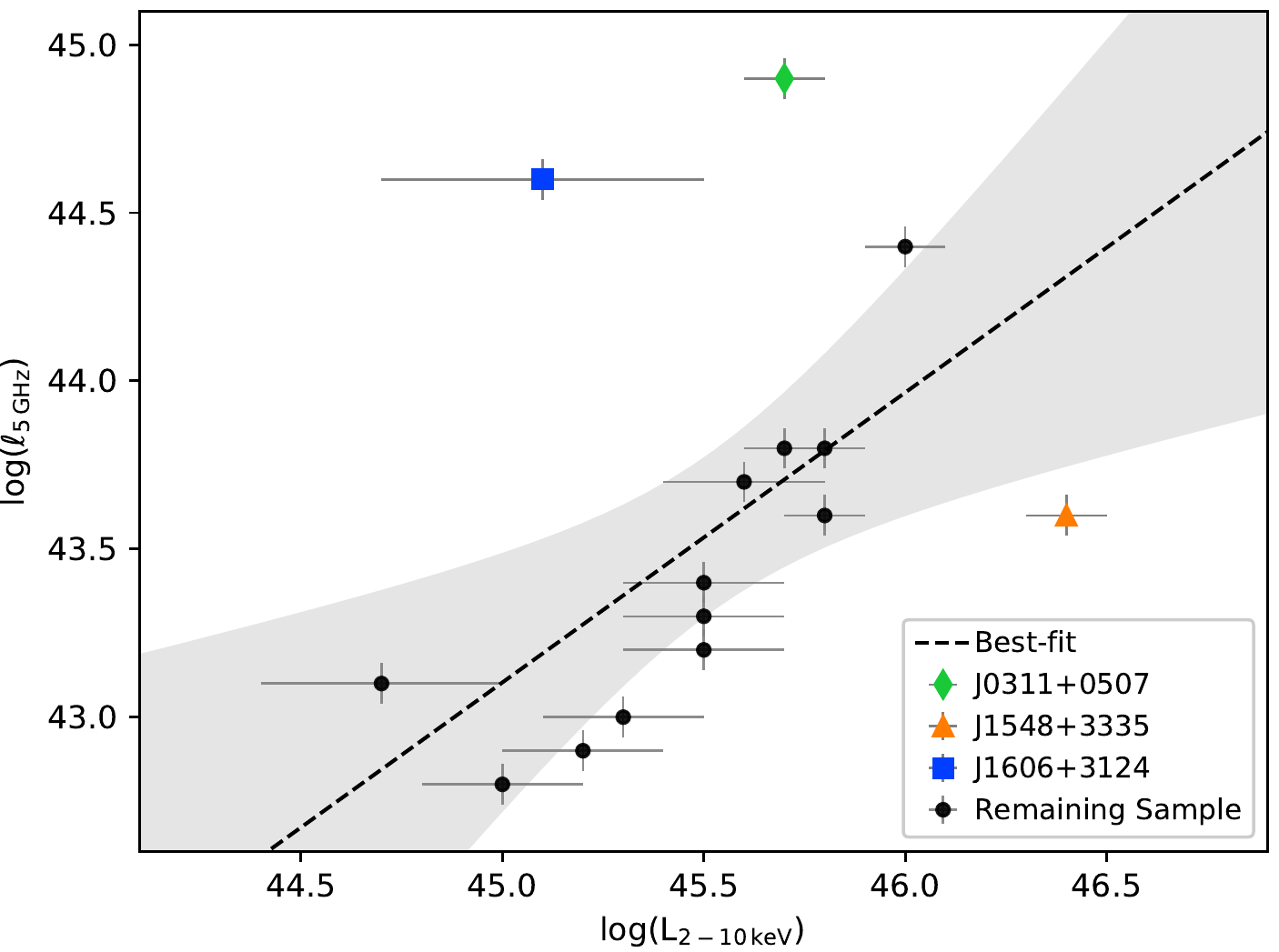} 
  \end{tightcenter}
\caption{Comparison of rest-frame radio ($\ell_{\rm5\,GHz}$) and X-ray ($L_{\rm 2-10\,keV}$) luminosities. The black dotted line is the best-fit relation, while the grey region is the $1\sigma$ confidence level.
Radio--X-ray ratios broadly agree for the majority of the sample. Three sources (J0311+0507, J1548+3335, and J1606+3124) are greater than $1\sigma$ from the best fit.}
\label{fig:radio}
\end{figure}

Recent radio and X-ray studies of 24 young radio quasars at $z < 1$ by \cite{Sobolewska2019} identified 29\% of their sample to have intrinsic hydrogen column densities $N_{\rm H}^i > 10^{23} \rm\,cm^{?2}$. A comparison of obscured quasar populations over a broad redshift range may place constraints on the evolution of obscured sources as well as explore the environmental conditions in the early universe. Thus, we compared the obscured population percentage of the low-redshift quasars to results from our high-redshift sample. 

Based on the X-ray spectral analysis in Section~\ref{sect:lumin}, only one source in our high-redshift sample at $4.5 < z < 5.0$, or 7\%, is potentially X-ray obscured. No evidence of intrinsic absorption was detected in the remaining 14 sources at a $3\sigma$ upper bound of $1.1 \times 10^{23}\rm\ cm^2$. We stress that the observed difference in the obscured population percentage between low- and high-redshift is at a marginal significance due to the small sample size, but this difference may indicate that a larger proportion of obscured sources is present at low-redshift. It is therefore possible that the dense environmental conditions required for X-ray obscured quasars require time to develop. Future observations of young radio quasars over broader redshifts will further constrain this parameter range and consequently test the dependence on environmental conditions for these objects.

\section{Outliers within the Sample}
\label{sect:outlier}

\begin{figure}
  \begin{tightcenter}   
    \includegraphics[width=0.99\linewidth]{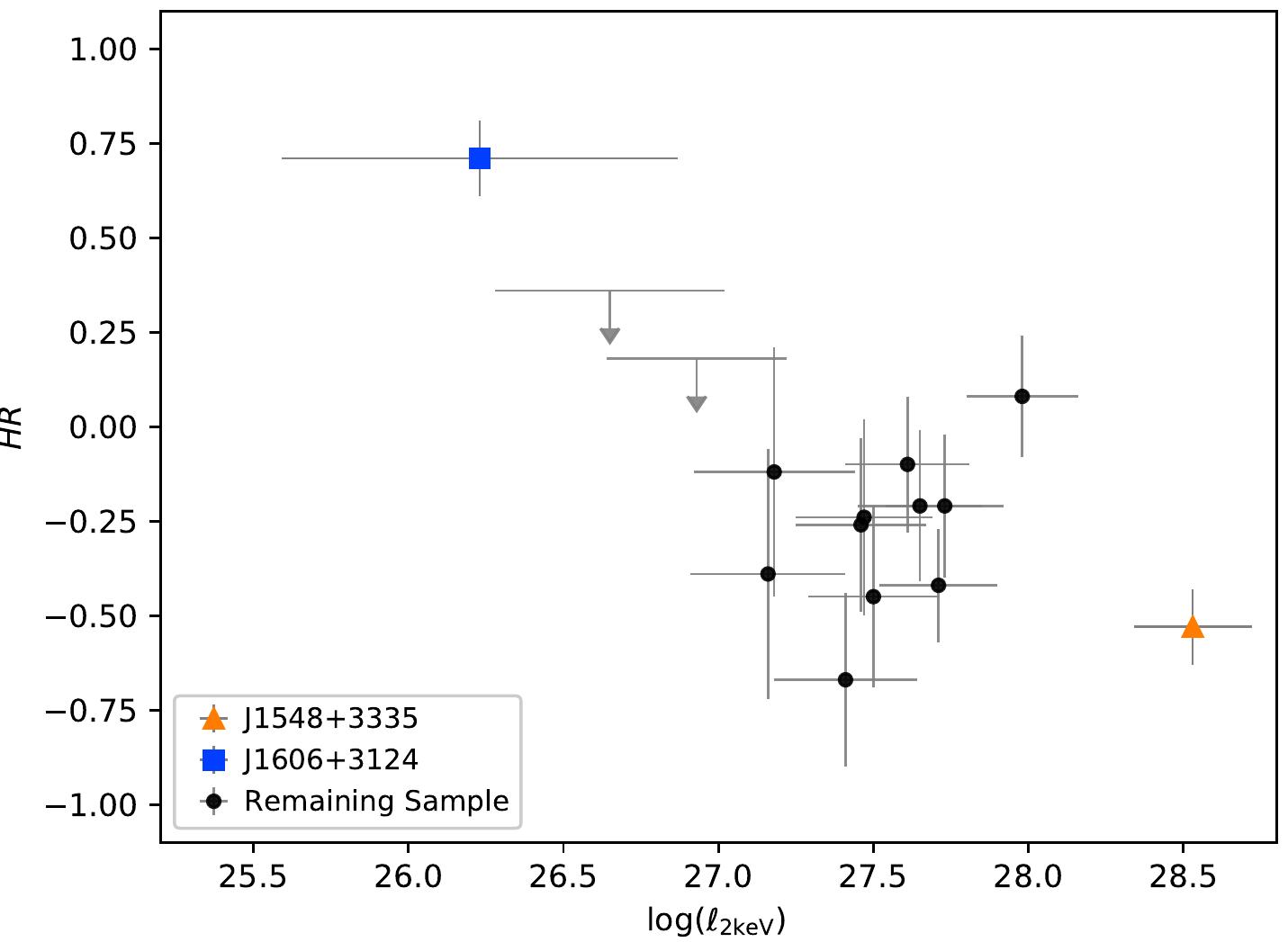}  
  \end{tightcenter}
\caption{Comparison of Hardness Ratio ($HR$) versus 2\,keV luminosity $\ell_{\rm 2\,keV}$ for the sample. The sources J1548$+$3335 ($z=4.68$) and J1606$+$3124 ($z= 4.56$) are clear outliers from the remainder of the sample due to their X-ray luminosities and $HR$.}
\label{fig:hr}
\end{figure}

\subsection{High X-Ray Luminosity Source J1548$+$3335} 
\label{sect:J1548}

Amongst the sample, J1548$+$3335 stands out as the brightest X-ray source with a 2\,keV luminosity $\ell_{\rm 2\,keV} = 3.4 \times 10^{28}\rm\,erg\,\,s^{-1}\,Hz^{-1}$, over 10 times greater than other sources at similar redshift (Figure~\ref{fig:hr}). It additionally has the highest $\alphaox$ in our analysis of $-1.24\substack{+0.07\\-0.08}$. The high X-ray emission from J1548 makes it an outlier within our sample, necessitating further investigation.

The high 2\,keV luminosity from J1548 is primarily attributed to its large X-ray photon index $\Gamma = 2.17\substack{+0.13\\-0.11}$, making it the softest X-ray source in the sample. The elevated soft X-ray emission may indicate the presence of an 6.4\,keV Fe\,K$\alpha$ emission line, which would be at an observed energy 1.13\,keV for $z=4.68$. Using the routine {\tt plot\_pvalue} in Sherpa, we added an Fe\,K$\alpha$ line to the spectral model from Section~\ref{sect:xray} and subsequently tested for a statistical improvement in the fit. The energy width of the line was allowed to freely vary. A $p$-value of 0.91 was obtained, so we therefore cannot discern a statistical improvement from the inclusion of the Fe\,K$\alpha$ line. 

As an explanation of the high X-ray luminosity from J1548 is limited by the available snapshot data, further analysis requires follow-up, deep-exposure X-ray observations of the system. A detection of the Fe\,K$\alpha$ line strength would constrain the Fe\,K reflection region and the iron abundance of the young source \citep{Balokovi2018}, while detection of line broadening may be used to estimate Compton scattering rates in the hot disk atmosphere \citep{Reynolds2014}. Multiepoch data from additional observations may also be used to constrain the average intensity from J1548 and better characterize its $\alphaox$ relative to the remaining sample. 

\subsection{Compton-Thick Obscuration in J1606+3124}
\label{sect:J1606}

As discussed in Section~\ref{sect:hr}, J1606$+$3124 has a hardness ratio of  $+0.71\substack{+0.15\\-0.10}$ that deviates from the sample average of  $-0.29\substack{+0.15\\-0.21}$ by more than $3\sigma$. Additionally, our spectral analysis of J1606 in Section~\ref{sect:lumin} found its spectrum to be abnormally flat with a 2\,keV luminosity $\ell_{\rm 2\,keV} = 0.2\times 10^{27}\rm\,erg\,\,s^{-1}\,Hz^{-1}$, well-below the remaining sample (Figure~\ref{fig:hr}). These properties are consistent with a dense absorption column at the source. 

Analysis of the observed spectrum for J1606 in Section~\ref{sect:xray} found  $N_{\rm H}^{i} = 1.34\substack{+0.45\\-0.38} \times 10^{24}\rm\,cm^{-2}$ when $\Gamma$ was fixed to the best-fit value from the remaining sample. Both $N_{\rm H}$ and $\Gamma$ parameters could not be simultaneously fit due the low number of source counts. While this spectral fit cannot conclusively determine the physical properties of the source, it does show that J1606 may be adequately modeled as a Compton-thick\footnote{A Compton-thick quasar is one where its intrinsic column density exceeds the inverse of the Thomson cross-section \mbox{($N_{\rm H}^{i} \geq  1.5\times10^{24}\rm\,cm^{-2}$)}, resulting in high obscuration of the soft X-ray band.} (CT) source. 

Further verification that J1606 is CT may be done by comparing its spectral properties with known CT sources. As high-redshift CT sources are rare  \citep{Vito2019b}, we compared J1606 with the nearby CT source OQ$+$208 ($z$ = 0.077, $N_{\rm H}^i = 10^{23}-10^{24}\rm\,cm^{-2}$) which has been exquisitely modeled over the 0.5--30.0\,keV rest-frame energy band \citep{Sobolewska2019b}. As we lack the counts to fit J1606 with similarly complex models as OQ$+$208, we fit both systems with a {\tt phabs$\cdot$powerlaw} expression over the rest-frame 0.5--7.0\,keV band and compared $\Gamma$ between the quasars. Fitting OQ$+$208 and fixing its Galactic absorption, $\Gamma$ was found to be $1.09 \pm 0.06$. Fitting J1606 with the same model gives $\Gamma = -0.11\substack{+0.41\\-0.43}$. We note that $\Gamma < 1.0$ is rare for an unabsorbed source, even in cases of relativistic beaming \citep{Ighina2019}. Thus, J1606 is likely heavily obscured, even in comparison to the CT source OQ$+$208. 

Altogether, it is highly probable that J1606 is a heavily obscured CT source. Additional X-ray observations of J1606 are required to accurately model the intrinsic column density, and deep-exposure observations will allow for simultaneous fits of both the column density and photon index. 

\subsection{High Radio--X-Ray Luminosity Ratio from J0311$+$0507}
\label{sect:J0311}

The comparison of radio and X-ray luminosities in Figure~\ref{fig:radio} shows two outliers with above-average radio--X-ray luminosity ratios, J1606$+$3124 and J0311$+$0507. The X-ray deficit in J1606 is likely attributed to the intrinsic absorption at the source, as examined in Section~\ref{sect:J1606}. However, there is no evidence of intrinsic absorption in J0311, necessitating further analysis. 

Previous 1.7 and 5\,GHz observations with both the Multi-Element Radio Linked Interferometer Network (MERLIN) and European Very Long Baseline Interferometer Network (e-EVN) indicated J0311 to have an FR\,II structure with a high degree of asymmetry and a linear scale of 18.7\,kpc, or 2\farcs8 \citep{Parijskij2014}. X-ray radiation will accompany this extended structure via shocked ISM emission and nonthermal emission from the jet and lobes \citep{Heinz1998,Reynolds2001}. At the distance scale of 2\farcs8, \chandra\ is capable of resolving extended features, if present, in J0311. However, our analysis in Section~\ref{sect:extend} found no evidence of extended emission from the source greater than the telescope's spatial resolution of limit 0\farcs5. Thus, the observed X-rays are localized to the core of J0311.

As noted in Section~\ref{sect:radio}, the reported radio luminosities represent the total emission for each source.  Taking the ratio of the core to the total radio flux from \cite{Parijskij2014}, we estimated the rest-frame 5\,GHz for the core of J0311 to be $\ell_{\rm 5\,GHz} = 12.6\times 10^{33} \rm\,erg\,s^{-1}\,Hz^{-1}$. This value gives a radio--X-ray luminosity ratio for J0311 that is within $1\sigma$ of the best fit in Figure~\ref{fig:radio}. Thus, the inclusion of the extended radio emission in our initial analysis explains the anomalously high radio--X-ray luminosity ratio as compared to the remaining sample. Follow-up \chandra\ observations of J0311 at deeper exposures are required to potentially resolve the extended X-ray counterpart in this high-redshift FR\,II source. 

\section{Conclusions}
\label{sect:conclude}

We analyzed \chandra\ observations of 15 young radio quasars at redshift $4.5 < z < 5.0$ to determine X-ray properties of this distant quasar population. All sources were successfully detected in the 0.5--7.0\,keV energy band, and X-ray emission spectra were extracted. Fitting each spectrum with a power-law model and allowing the photon index $\Gamma$ to freely vary, the measured $\Gamma$ ranged between $[-0.11,2.17]$ for the sample, where the simultaneous best-fit photon index was $\Gamma = 1.5\pm0.1$. Unabsorbed rest-frame 2--10\,keV luminosities were measured to be between $[0.5,23.2] \times 10^{45}\rm\,erg\ s^{-1}$ for the sample. Radial profiles were extracted from each observation to investigate for extended X-ray emission, but no X-ray extension was observed for any target.

We determined the optical--X-ray power-law slope $\alphaox$ for each source using optical/UV data available in the literature. The measured $\alphaox$ values ranged between [-1.24,-2.08]. Results from this work were compared against other X-ray bright quasar surveys. An observed anticorrelation trend between  $\alphaox$\,--\,$\ell_{2500\angstrom}$ was measured and shown to agree well with independent estimates \citep[e.g.,][]{Nanni2017}. We additionally measured radio--X-ray luminosity ratios for our sources, and the results were broadly consistent with other quasar surveys regardless of radio-loudness and redshift. These multiwavelength results reinforce the lack of spectral evolution for quasars over a broad redshift range.

We identified three significant outliers from the sample based on their X-ray properties and multiwavelength relationship. J1548$+$3335 stands out as the brightest X-ray source with a 2--10\,keV luminosity of $23.2 \times 10^{45}\rm\,erg\ s^{-1}$, a factor of $\sim$\,10 larger than the next brightest source at a similar redshift. J0311$+$0507 was verified to have an FR\,II structure at a linear scale of 2\farcs8 in radio, while no extended X-ray emission was observed above the 0\farcs5 spatial limit of \chandra. Spectral modeling of J1606$+$3124 indicates that it has an intrinsic column density greater than $10^{24} \rm\,cm^{-2}$ and is therefore Compton-thick. Follow-up \chandra\ observations at deeper exposures are recommended for these outliers in order to better quantify their unique properties and to place them in context with the young radio quasar population.    

\acknowledgements{
Support for this work was provided by the National Aeronautics and Space Administration through \chandra{} Award Numbers GO8-19093X \& GO0-21101X issued by the Chandra X-ray Observatory Center, which is operated by the Smithsonian Astrophysical Observatory for and on behalf of the National Aeronautics Space Administration under contract NAS8-03060.  A.S., M.S., and D.A.S. were supported by NASA contract NAS8-03060 (Chandra X-ray Center). Work by C.C.C. at the Naval Research Laboratory is supported by NASA DPR S-15633-Y. \L.S. was supported by the Polish NSC grant 2016/22/E/ST9/00061.
\software{\ciao\ v4.11.5, \caldb\ v4.8.5 \citep{Fruscione2006}, Sherpa \citep{Freeman2001}, {\tt scipy} \citep{Virtanen2020}, {\tt linmix\_err} \citep{Kelly2007b} }
}

\bibliographystyle{aasjournal}
\bibliography{all_data}

\end{document}